# Improvements on INRIM Coaxial Microcalorimeter and Outcome of a Model Comparison

Luciano Brunetti, Luca Oberto, Marco Sellone, Nosherwan Shoaib and Emil Vremera, *Member, IEEE*

*Abstract*—This paper describes hardware and software improvements of the INRIM coaxial microcalorimeter together with their outcome on the primary power standard realization in the frequency band 0.05 – 40 GHz. A better temperature and power stabilization turned out to provide an improved signal/noise ratio and a drift reduction in every working condition of the microcalorimeter. The INRIM correction model is also compared to a traditional, but faster, one in terms of measurement uncertainty.

Outcomes are presented in form of a 2.92 mm thermoelectric power sensor calibration together with results that show the improved stability and repeatability of the measurement system.

## I. Introduction

IN the Radio Frequency (RF) and Microwave (MW) range, a key quantity always well defined and measurable is the electromagnetic power [1]. Therefore the power standard is of the utmost importance for primary electromagnetic metrology. All National Metrology Institutes (NMIs) realize the high frequency (HF) primary power standard, tracing the calibration of a thermal detector to the *dc power standard*. The *principle of equivalence of the thermal effects* is applied for that purpose. This technique has been introduced in the late 1950s and today it is usually referenced as microcalorimeter technique [1]-[3]. Up to now, alternatives do not exist yet, therefore the continuous improvement of microcalorimeter systems in terms of both hardware and software is very important for all NMIs. Even though microcalorimeters exist both in waveguide and in coaxial line with different performances [4]-[13], INRIM mainly developed coaxial

Manuscript received August xx, 2014.
L. Brunetti, L. Oberto, M. Sellone and N. Shoaib are with the Electromagnetism Division of the Istituto Nazionale di Ricerca Metrologica (INRIM), 10135 Torino, Italy (email: l.brunetti@inrim.it, l.oberto@inrim.it, m.sellone@inrim.it, n.shoaib@inrim.it).
E.Vremera is with the Faculty of Electrical Engineering of Gheorghe Agachi Technical University of Iasi, 700050 Iasi, Romania (email: evremera@ee.tuiasi.ro).

systems because of their broadband characteristics. Furthermore, it has been one of the first NMIs to propose the microcalorimeter based on the thermoelectric detection as an alternative to the more classical bolometric detection [14]-[24]. Sensors based on thermoelectric principle are less sensitive to ambient temperature variations, and are not downward frequency limited.

The improvements to INRIM measurement system reached recently a new level by refining temperature and power controls. The paper shows the effects of these improvements on the calibration of a coaxial thermoelectric power sensor in the frequency band 0.05 – 40 GHz, as already anticipated in [25].

## II. System Details

The INRIM microcalorimeter is an adiabatic dry microcalorimeter fitted with 2.92 mm twin coaxial-line inset. The system architecture is slightly different from that of the model cited in the literature of the same authors. It has been specifically designed to calibrate thermoelectric power sensors in terms of effective efficiency in the frequency range 0.01 – 40 GHz. The temperature stabilization of the microcalorimeter load is obtained by means of a combination of passive and active metal shields separated by polymeric foam as insulating material. The temperature control system is based on Peltier elements and a wire heater driven by PID controllers. It requires to operate inside a preconditioned room at the temperature of (23.0 ± 0.3) °C and relative humidity of (50 ± 5) %.

Former systems [14]–[24], placed inside the same preconditioned room, were able to maintain the measurement chamber at (25.00 ± 0.01) °C for about 50 minutes. In the new design the thermal stability has been increased of about one order of magnitude (about 3 mK) for a longer duration (more than 20 hours). This allows better measurement uncertainties when the microcalorimeter operates in critical conditions, that is, when the sensor losses are very low.

However, sensitivity and accuracy of the INRIM coaxial microcalorimeter have been improved not only with respect to its thermal control system, but also to the stabilization of the measurement power level.

To be more specific on the new design and with reference to Fig. 1, we modified the insulating sections and, furthermore, an external one (I IS) has been added to the thermostat to improve the thermal insulation of the microcalorimeter load against the external environment. Secondly, an additional passive planar shield (IV SHIELD) has been placed in front of the measurement port to reduce the thermal offset between reference and measurement channel even without power injection. Finally a sensitive temperature control has been applied to the massive aluminum cylinder (III SHIELD) that embraces the measurement chamber.

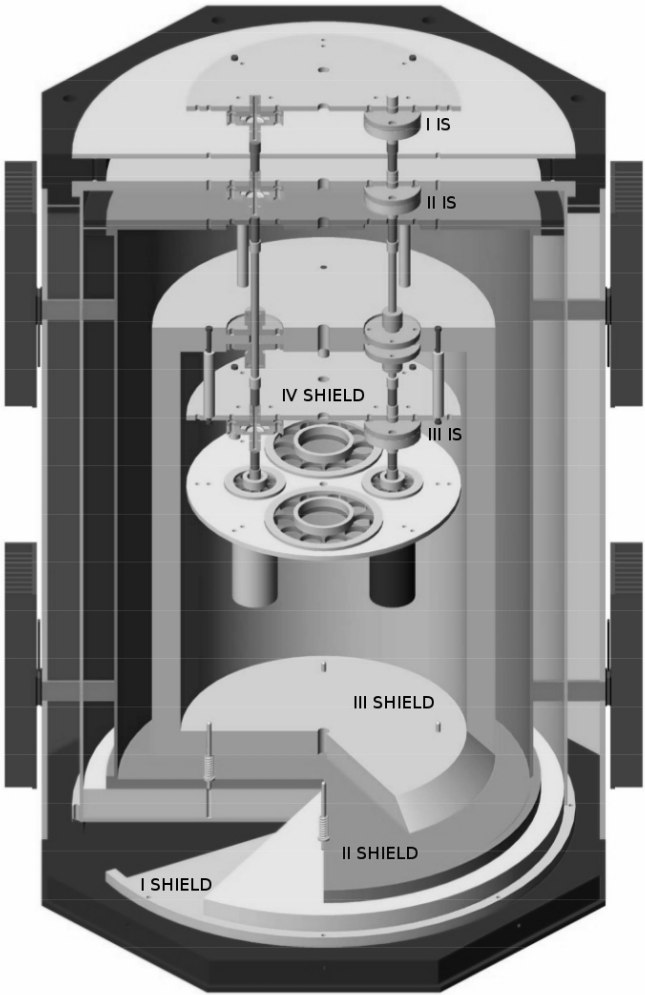

Fig. 1.   CAD picture shows the latest INRIM microcalorimeter that consists of the basic structure of the thermostat together with the inset elements.

The advantages of this hardware improvement on the temperature stability can be seen in Fig. 2 that shows the temperature behaviour in the microcalorimeter at the level of the thermopile fixture during the substitution of the reference power (REF) level (1mW at 1 kHz) with an equivalent HF power. The plot shows clearly that the temperature fluctuations inside the microcalorimeter are always below

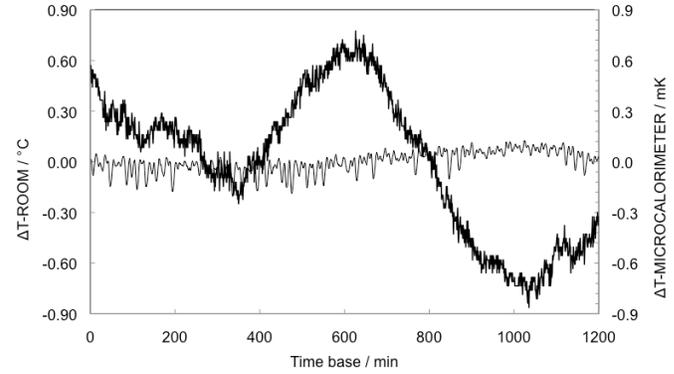

Fig. 2.   Temperature variations outside and inside microcalorimeter, during a typical measurement cycle. Left Y-axis reports the ambient temperature variations (thin line); right Y-axis shows the temperature fluctuations inside microcalorimeter (bold line).

1mK, three orders lower than the fluctuation of the environment.

Furthermore, a good thermal decoupling is confirmed by the calculated correlation coefficient between the external temperature variations and the thermopile output that results in 0.016. This coefficient has been evaluated by means of repeated measurements of both the temperature and the asymptotic value of the thermopile voltage and accoding to [26]-[27].

Another improvement concerns the power output stabilization of the generators used to perform the REF-HF-REF power substitution into the system. This has been obtained by adding an algorithm based on PID controllers and $\Sigma\Delta$-modulators to the measurement software. The improvement has a direct effect on the repeatability of the measurements.

### III.   Microcalorimeter Mathematical Model

The new coaxial microcalorimeter measures the *effective efficiency* $\eta_e$ of a thermoelectric sensor mount, which is defined as ratio of the measured power $P_M$, that is, the HF power actually converted into a dc output by the sensor, to the total absorbed power $P_A = (P_M + P_X)$:

$$\eta_e = \frac{P_M}{P_M + P_X},\tag{1}$$

where $P_X$ is the power loss in the sensor mount [15]-[17].

Operatively the expected value of $\eta_e$ can be obtained through a mathematical model that has been widely described in the literature by the same authors [15]-[24]:

$$\eta_e = \frac{e_2}{e_1 - \left(1 + |\Gamma_S|^2\right)\frac{e_{1SC}}{2}},\tag{2}$$

where $e_1$ and $e_2$ are the responses of the electrical thermometer of the microcalorimeter (i.e. a thermopile) to the HF power and to the REF power substituted into the system, respectively. The voltage $e_{1SC}$ corrects the microcalorimeter loss effects that result as dominant error contribution in the whole process of the power standard realization. This voltage is determined by means of the short circuit technique [15], [20], and it has to be halved to take into account the power reflected back by the short circuit. Finally, the term $(1+|\Gamma_S|^2)$ is an additional correction necessary to enhance the accuracy of the power standard when the reflection coefficient $\Gamma_S$ of the power sensor under calibration is not negligible [18].

We strongly support the use of model (2) but, since it requires the repetition of the whole measurement procedure twice to calibrate both the sensor and the microcalorimeter in short circuit condition, it is very time consuming.

Anyway we will demostrate, later in this paper, that the traditional method used in the past by INRIM and other NMIs, is less accurate. This traditional model can be derived directly from the $\eta_e$ definition (1) by adding the rate $\delta P_L$ of the microcalorimeter feeding line losses that influence $\eta_e$. It is given by:

$$\eta_e = \frac{P_M}{P_M + P_X + \delta P_L} = \frac{\eta_e^{raw}}{1 - \eta_e^{raw}\left(\dfrac{\delta P_L}{P_M}\right)} = \frac{e_2}{e_1 - e_2\left(\dfrac{\delta P_L}{P_M}\right)}, \quad (3)$$

where $\eta_e^{raw}$ is equal to $e_2/e_1$ and represents the uncorrected effective efficiency obtained from the measurements.

From the S-parameter theory, the perturbation term in the denominator of (3) can be expressed as a function of the feeding line transmission parameter $S_{12}$ and the power sensor reflection coefficient $\Gamma_S$. Then, under the reasonable hypothesis that only 50% of the feeding line losses influences the measurements [1], we obtain the following model:

$$\eta_e = \left(\frac{1+|S_{12}|^2\left(1-2|\Gamma_S|^2\right)}{2|S_{12}|^2\left(1-|\Gamma_S|^2\right)}\right)\eta_e^{raw}. \quad (4)$$

The previous hypothesis about the influence of line losses can be justified considering the thermodynamic model of a line section with uniformly distributed losses. If its ports are at the same temperature then half of the generated thermal energy leaves the line through each port.

Model (4) avoids running the microcalorimeter in short circuit condition, but has intrinsic limitations that we will highlight in the next paragraph, where we will show how its performance relates to the model (2).

## IV. MEASUREMENTS AND DATA ANALYSIS

Experimental work consists of the calibration of a thermoelectric power sensor fitted with a 2.92 mm connector so to realize the primary power standard in the frequency band 0.01 – 40 GHz.

Measurements have been performed at 1 GHz step, but numerical value of the measurand is hereby given, together with its uncertainty, only at seven specific frequencies (50 MHz, and 1, 10, 18, 26.5, 33, 40 GHz), known to be critical or limiting for some coaxial connector/line types.

Examples of detailed uncertainty budgets are given to support comments and conclusion. Raw effective efficiency $\eta_e^{raw}$ has been calculated by means of fitting/averaging processes applied to thermopile outputs $e_1$ and $e_2$, as already described in [16]-[24]. In short, we use a fitting method based on the Levenberg-Marquardt algorithm that requires as data input the thermopile voltage and the time base value together with their uncertainties [26]. The algorithm returns the asymptotic value of the thermopile output. Fig. 3 and its closeup Fig. 4 show the output of the mentioned calculation process superimposed to the real thermopile output voltage, for one complete REF-HF-REF substitution cycle at 40 GHz. Applying the mentioned process to several power substitution cycles, we can obtain the expected mean value of the measurand with an associated standard deviation ($\sigma$).

Correction terms appearing in (2) and (4) are calculated by means of the software mentioned before, when it is the case ($e_{1SC}$) and through independent measurements of reflection coefficient $\Gamma_S$ and transmission parameter $S_{12}$.

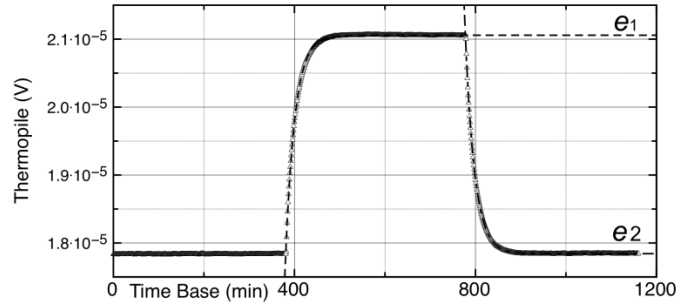

Fig. 3. Thermopile response at 1 mW REF-HF-REF power substitution, together with the fitting results.

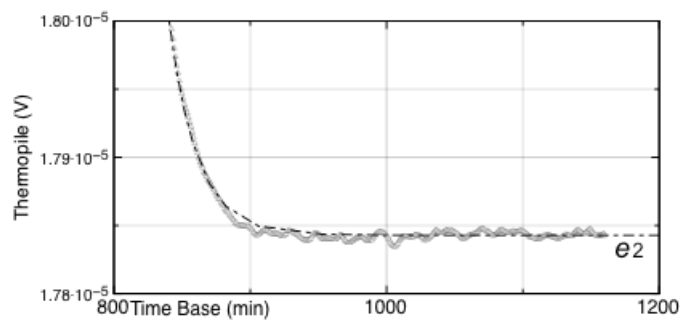

Fig. 4. Expansion of the thermopile response at 1 mW HF-REF substitution step to better highlight the quality of the fitting process.

Table I shows the raw effective efficiency and the values of the same measurand corrected according to both models (2) and (4), together with their uncertainty terms, at the specific frequencies. The total measurement uncertainty of the measurand $\eta_e$ has been calculated by applying the Gaussian error propagation on (2) and (4), and considering the possible existence of correlation among the influence quantities as suggested in [27].

Table II shows the detailed uncertainty budgets related to models (2) and (4) at 40 GHz. Correlation terms exist only for the voltages $e_1$ and $e_2$, but they are not reported in Table I and II because their contribution to the uncertainty turned out to be negligible if compared to other uncertainty contributions. No correlation exists among the quantities $e_{1SC}$, $\Gamma_S$, $S_{12}$, $e_1$ and $e_2$, because they are either measured with different independent methods or at different times and conditions.

At first glance, the results of Table I show a good agreement between the expected values of effective efficiency obtained by using model (2) and (4), even though at all frequencies, model (4) results in bigger uncertainties. This behaviour was however expected, because there are difficulties in determining the actual values of both the transmission parameter $S_{12}$ and the HF power loss rate *on site*, without dismounting the line inset. We considered that only 50% of the losses of the last insulating section (III IS) affects the load, as descibed in Sec. III. This hypothesis revealed to be reasonable, but evidently not enough valid to obtain the best expected value of $\eta_e$ with the best uncertainty.

short circuit condition (model 2). Indeed, in this case the voltage $e_{1SC}$ automatically account for the loss rate of the feeding line, whatever long and complex it is. Of course, this turns out to be a benefit for the total measurement accuracy.

Looking at Table II, we see that the most limiting factor in the accuracy budget is the term $S_{12}$. At present it is very difficult to find the actual values of the feeding line losses *on site*. This condition implies to be very conservative with both its value and uncertainty. Furthermore and unfortunately, the sensitivty coefficient $c(S_{12})$ derived from the mathematical model (4) is quite high; this made the uncertainty contribution worse.

## V. CONCLUSION

After having introduced thermodynamic improvements to INRIM coaxial microcalorimeter, we performed a full calibration of a thermoelectric power sensor in the frequency band $0.01 - 40$ GHz by using two different correction models. The outcome of this particular comparison, confirms that the actual INRIM microcalorimeter exhibits superior accuracy when it is calibrated by means of the short circuit technique. The correction technique based on the measurement of scattering parameters by means of network analyzer allows to save time, but gives as results a measurement accuracy significantly lower, at least for the INRIM system. In any case it indirectly confirms the validity of the model (2), that is the official correction currently applied to the coaxial microcalorimeter measurement at INRIM.

TABLE I
CALIBRATION LIST OF THERMOELECTRIC POWER STANDARD

| Freq. | $\eta_e^{raw}$ | $u(\eta_e^{raw})$ | $\eta_e$ | $u(\eta_e)$ | $\eta_e$ | $u(\eta_e)$ |
|---|---|---|---|---|---|---|
| (GHz) | | $1\sigma$ | Mod. (2) | $1\sigma$ | Mod. (4) | $1\sigma$ |
| 0.05 | 0.9879 | 0.00045 | 0.9923 | 0.00049 | 0.9926 | 0.0054 |
| 1 | 0.9662 | 0.00043 | 0.9806 | 0.00048 | 0.9804 | 0.0057 |
| 10 | 0.9097 | 0.00038 | 0.9383 | 0.00044 | 0.9371 | 0.0128 |
| 18 | 0.8855 | 0.00037 | 0.9270 | 0.00043 | 0.9281 | 0.0128 |
| 26.5 | 0.8725 | 0.00036 | 0.9154 | 0.00043 | 0.9118 | 0.0356 |
| 33 | 0.8633 | 0.00035 | 0.9035 | 0.00043 | 0.9031 | 0.0373 |
| 40 | 0.8473 | 0.00034 | 0.8974 | 0.00043 | 0.8980 | 0.0307 |

TABLE II
DETAILS OF UNCERTAINTY BUDGET AT 40 GHZ FOR THERMOELECTRIC STANDARDS (EXCLUDING ADIMENSIONAL REFLECTION AND TRANSMISSION COEFFICIENTS; QUANTITIES AND RELATED UNCERTAINTIES ARE IN VOLT)

| Influence Variable | Measured Value | Measurement Uncertainty | Sensitivity coefficient | Uncertainty Contribution |
|---|---|---|---|---|
| | $y$ | $u(y)$ | $\|c(y)\|$ | $c(y)u(y)$ |
| *Correction model (2)* | | | | |
| $e_1$ | 2.1058E-05 | 7.0E-09 | 4.5138E+04 | 0.00031 |
| $e_2$ | 1.7843E-05 | 4.0E-09 | 5.0297E+04 | 0.00020 |
| $e_{1SC}$ | 2.3273E-06 | 7.3E-09 | 2.2800E+04 | 0.00016 |
| $\Gamma_S$ | 0.1013 | 0.0130 | 0.0106 | 0.00013 |
| $u(\eta_e)$ | | | | 0.00043 |
| *Correction model (4)* | | | | |
| $e_1$ | 2.1058E-05 | 7.0E-09 | 4.5138E+04 | 0.00031 |
| $e_2$ | 1.7843E-05 | 4.0E-09 | 5.0297E+04 | 0.00020 |
| $S_{12}$ | 0.6992 | 0.0122 | 2.5084 | 0.03058 |
| $\Gamma_S$ | 0.1013 | 0.0130 | 0.1874 | 0.00243 |
| $u(\eta_e)$ | | | | 0.03068 |

The mentioned assumptions are not requested if we introduce the correction based on the method that uses the

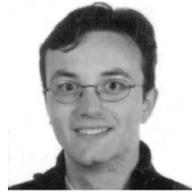

**Luca Oberto** was born in Pinerolo (Torino), Italy, on June 9, 1975. He received the M.S. degree in Physics from the University of Torino in 2003 and the Ph.D. in Metrology from the Politecnico di Torino in 2008.

From 2002 to 2003 he was with the Istituto Nazionale di Fisica Nucleare (INFN), Torino Section, working at the COMPASS experiment at CERN, Geneva, Switzerland. From 2003 he is with the Istituto Nazionale di Ricerca Metrologica (INRIM), Torino, Italy. His research interests are in the field of high frequency and THz metrology and in the realization and characterization of superconductor-insulator-superconductor mixers for astrophysical applications in the millimeter- and sub millimeter-wave domain.

Dr. Oberto is member of the Associazione Italiana Gruppo di Misure Elettriche ed Elettroniche (GMEE). He was recipient of the 2008 Conference on Precision Electromagnetic Measurements Early Career Award and of the GMEE 2010 "Carlo Offelli" Ph.D. prize.

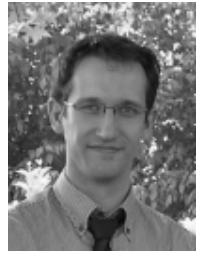

**Marco Sellone** was born in Turin, Italy, on September 25, 1979. He received the M.S. degree in Physics from the University of Torino, Italy, in 2004 and the Ph.D. degree in Metrology from the Politecnico di Torino in 2009.

Since 2005, he has been with the Istituto Nazionale di Ricerca Metrologica (INRIM), Torino, Italy. His research interests are the High Frequency Metrology mainly regarding VNAs measurements and uncertainty evaluation, and novel VNA measureents applications on materials.

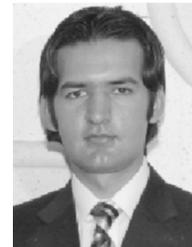

**Nosherwan Shoaib** was born in Pakistan on October 7, 1986. He has completed the Bachelors of Communications Systems Engineering from Institute of Space Technology, Islamabad, Pakistan in August 2008. He received the Masters of Electronics Engineering degree from the Politecnico Di torino, Italy in September 2011. After graduation, he served as a Lecturer at Heavy Industries Taxila Education Complex (HITEC) University, Pakistan.

Since 2012, he has been a Doctorate Student in Metrology at Politecnico di Torino, Italy. He is currently working on high frequency measurement techniques and measurement uncertainty evaluation at Istituto Nazionale di Ricerca Metrologica (INRIM), Torino, Italy. He has been awarded the ARFTG PhD Student Sponsorship Award 2013.

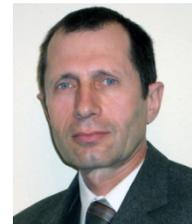

**Emil T. Vremera** was born in Romani-Neamt, Romania, on September 11, 1953. He received the M.Sc. degree in electronics and the Ph.D. degree in electrical measurements from "Gheorghe Asachi" Technical University of Iasi, Romania, in 1977 and 1998, respectively.

He is currently with the Department of Electric Measurements, Faculty of Electrical Engineering, "Gheorghe Asachi" Technical University of Iasi, where he joined in 1984 first as an Assistant Professor and then as a Professor. He teaches electric and electronic measurements for the students in the electronic area. Since 2001, he has been developing a research activity on RF power measurements. He is also an Associated Scientist with Istituto Nazionale di Ricerca Metrologica, Turin, Italy. His main research interests concern measurement techniques of the electric and magnetic quantities, analog to digital conversion for second-order quantities, and virtual instrumentation.

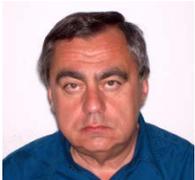

**Luciano Brunetti** was born in Asti, Italy, on September 11, 1951. He received the M.S. degree in Physics from the University of Torino, Italy, in 1977.

Since 1977 he has been working at Istituto Nazionale di Ricerca Metrologica (INRIM, formerly IEN "Galileo Ferraris"), Torino, Italy. He has been dealing both with theoretical and experimental research in the field of high frequency primary metrology. His main task has always been the realization and the dissemination of the national standard of power, impedance and attenuation in the microwave range. In the last years he has been involved in the design and characterization of millimeter and microwave devices working at cryogenic temperature; he collaborates also at the characterization of complex magnetic alloys at high frequency. Actually he is taking care of the extension of the national electrical standards in the millimeter wavelength range.